\documentclass{PoS}

\usepackage{graphicx}
\usepackage{amsmath}
\usepackage{amssymb}

\newcommand{\be}{\begin{equation}}
\newcommand{\ee}{\end{equation}}
\newcommand{\bea}{\begin{eqnarray}}
\newcommand{\eea}{\end{eqnarray}}
\newcommand{\bi}{\begin{itemize}}
\newcommand{\ei}{\end{itemize}}
\newcommand{\ben}{\begin{enumerate}}
\newcommand{\een}{\end{enumerate}}
\newcommand{\bt}{\begin{tabbing}}
\newcommand{\et}{\end{tabbing}}

\newcommand{\dtau}{{\Delta \tau}}

\newcommand{\dHave}{{\langle \Delta H \rangle}}
\newcommand{\edHave}{{\langle e^{-\Delta H} \rangle}}
\newcommand{\Nmd}{{N_{\rm MD}}}
\newcommand{\Nmdref}{{N_{{\rm MD},P=0.80}}}
\newcommand{\Ninv}{{N_{\rm inv}}}
\newcommand{\Nsmr}{{N_{\rm smr}}}
\newcommand{\sgnZ}{{\epsilon_Z}}
\newcommand{\sgnP}{{\epsilon_p}}
\newcommand{\sgnM}{{\epsilon_M}}
\newcommand{\mres}{{m_{\rm res}}}

\title{
   \begin{picture}(0,0)(0,0)%
   \put(350,75){\makebox(0,0)[l]{\textnormal{\normalsize KEK-CP-293}}}%
   \end{picture}%
   Large-scale simulations with chiral symmetry
}

\ShortTitle{Large-scale simulations with chiral symmetry}

\author{
   JLQCD Collaboration: 
   \speaker{T.~Kaneko}$^{a,b}$\thanks{E-mail: takashi.kaneko@kek.jp}, 
   S.~Aoki$^{c}$, 
   G.~Cossu$^a$, 
   H.~Fukaya$^d$, 
   S.~Hashimoto$^{a,b}$
   and 
   J.~Noaki$^{a}$
   \\
   \\
   \\
   \llap{$^a$}
   High Energy Accelerator Research Organization (KEK),
   Ibaraki 305-0801, Japan 
   \\
   \llap{$^b$}
   School of High Energy Accelerator Science,
   The Graduate University for Advanced Studies (Sokendai),
   Ibaraki 305-0801, Japan
   \\ 
   \llap{$^c$}
   Yukawa Institute for Theoretical Physics,
   Kyoto University, 
   Kyoto 606-8502, Japan
   \\ 
   \llap{$^d$}
   Department of Physics, Osaka University, 
   Osaka 560-0043, Japan
}

\abstract{
We carry out a comparative study 
among five-dimensional formulations of chirally symmetric fermions 
about the algorithmic performance, 
chiral symmetry violation and topological tunneling
to find a computationally inexpensive formulation 
with good chiral symmetry.
With our choice of the lattice action, 
we have launched large-scale simulations 
on fine lattices
aiming at a precision study of light and heavy quark physics.
We report on the comparative study, 
current status of the large-scale simulations,
and preliminary results 
on the residual quark mass and auto-correlation.
}

\FullConference{
31st International Symposium on Lattice Field Theory - LATTICE 2013\\
July 29 - August 3, 2013\\
Mainz, Germany}

\begin{document}

\begin{table}[b]
\begin{center}
\caption{
   Simulation setup in our comparative study. 
   The first three columns show our choices of 
   the five-dimensional formulation: 
   the number of smearing $\Nsmr$, kernel operator $H_M$
   and sign function approximation $\sgnM$. 
   We also list simulation parameters, 
   namely $\beta$ and the bare quark mass in lattice units $am_{ud}$,
   as well as results for $a^{-1}$ and $M_\pi$.
}
\vspace{3mm}
\label{tbl:compara:param}
\begin{tabular}{lll|ll|ll}   \hline
   $\Nsmr$ & $H_M$  & $\sgnM$  & $\beta$  & $a^{-1}$ [GeV]  
                               & $am_{ud}$ & $M_{\pi}$ [MeV] 
   \\ \hline \hline
   0  & $H_W$  & $\sgnZ$ & 4.27 & 1.98(6)
                         & 0.0095,  0.0060,  0.0035  
                         & 463(17), 375(17), 346(25)
   \\ \hline
   0  & $H_T$  & $\sgnZ$  & 4.11 & 1.92(6)
                          & 0.0200,  0.0120,  0.0065 
                          & 543(18), 419(15), 318(15)
   \\ \hline
   0  & $H_T$  & $\sgnP$  & 4.11 & 1.97(5)
                          & 0.0200,  0.0090,  0.0040
                          & 623(19), 483(16), 400(15)
   \\ \hline
   0  & $2H_T$ & $\sgnP$  & 4.11 & 1.94(6)
                          & 0.0200,  0.0120,  0.0065 
                          & 554(18), 434(17), 356(16)
   \\ \hline \hline
   3  & $H_W$  & $\sgnZ$  & 4.29 & 1.94(6)
                          & 0.0145,  0.0090,  0.0050
                          & 472(18), 401(17), 330(17)
   \\ \hline
   3  & $H_T$  & $\sgnP$  & 4.18 & 2.00(8)
                          & 0.0250,  0.0170,  0.0090
                          & 534(23), 423(20), 374(23)
   \\ \hline
   3  & $2H_T$ & $\sgnP$  & 4.18 & 2.06(9)
                          & 0.0250,  0.0170,  0.0090
                          & 524(24), 469(24), 364(25)
   \\ \hline \hline
   6  & $2H_T$ & $\sgnP$  & 4.18 & 2.11(6)
                          & 0.0250,  0.0170,  0.0090
                          & 511(17), 430(16), 337(20)
   \\ \hline
\end{tabular}
\end{center}
\end{table}


\section{Introduction}

In the last several years, 
we performed an extensive study of QCD
vacuum and light hadron physics 
by using the overlap action which exactly preserves chiral symmetry
\cite{JLQCD:overlap}.  
Our next target is a precision study of heavy flavor physics
in collaboration with flavor factory experiments, 
such as the SuperKEKB / Belle II experiment,
for a stringent test of the Standard Model. 

Since the overlap action is computationally too expensive 
to simulate small lattice spacings $a \!\ll\! m_c^{-1}$
on reasonably large lattices,
we carried out a systematic comparative study of 
a class of five-dimensional formulations 
that approximately satisfy the Ginsparg-Wilson relation
to construct a computationally cheap formulation 
with good chiral symmetry. 
In this article,
we report on the comparative study and 
the status of the on-going large-scale simulations 
with our choice of the lattice action.


\section{Comparative study}

We test five-dimensional fermion formulations~\cite{Moebius}
in this comparative study.
The four-dimensional effective Dirac operator 
is given by
\bea
  \frac{1+m_q}{2} 
+ \frac{1-m_q}{2} \gamma_5\, \epsilon_M\left( H_M \right),
\label{eqn:compara:D4D}
\eea
where the Hermitian kernel operator $H_M$ and the approximation of 
its sign function $\epsilon_M$ can be chosen
by tuning parameters appearing in the five-dimensional Dirac operator.
Popular choices of $H_M$ are 
the Wilson kernel $H_W\!=\!\gamma_5 D_W$,
where $D_W$ is the Wilson-Dirac operator, 
for the overlap fermions,
and the Shamir kernel $H_T\!=\!\gamma_5 D_W / (2+D_W)$
for the standard domain-wall fermions. 
We also test a scaled Shamir kernel $2H_T$~\cite{Moebius}.
While  $2H_T$ has the same condition number as $H_T$,
its low-lying eigenvalues are scaled up by a factor of 2.
These kernels are combined with the Zolotarev ($\sgnZ$) 
and polar decomposition ($\sgnP$) approximations. 
By applying up to 6 level stout smearing~\cite{smearing} 
($\Nsmr\!=\!0,3,6$),
we test 8 different formulations listed 
in Table~\ref{tbl:compara:param}. 

We carry out numerical simulations of two-flavor QCD 
by using these formulations and the tree-level Symanzik gauge action
to study the performance of the Hybrid Monte Carlo (HMC) algorithm, 
chiral symmetry violation
and topological tunneling.
On a $16^3\!\times\!32$ lattice, 
we simulate three pion masses in the range of 
$300 \!\lesssim\! M_\pi \mbox{[MeV]} \!\lesssim\! 600$
at a single lattice cut-off around $a^{-1}\!\simeq\!2$~GeV.
The fifth dimensional size is set to $N_5\!=\!12$.
We set the range of the Zolotarev approximation $\sgnZ(x)$ 
to $x\!\in\![0.2,7.0]$ $([0.4,7.0])$ for 
$H_W$ without (with) smearing, and $[0.1,1.5]$ for $H_T$.
Our statistics are 1,000 trajectories in each simulation.
Parameters and results for $a^{-1}$ and $M_\pi$ 
are summarized in Table~\ref{tbl:compara:param},
where $r_0\!=\!0.462(11)(4)$~fm~\cite{r0:MILC} is used as input to fix $a$.  

\begin{figure}[t]
\begin{center}
   \includegraphics[angle=0,width=0.47\linewidth,clip]%
                   {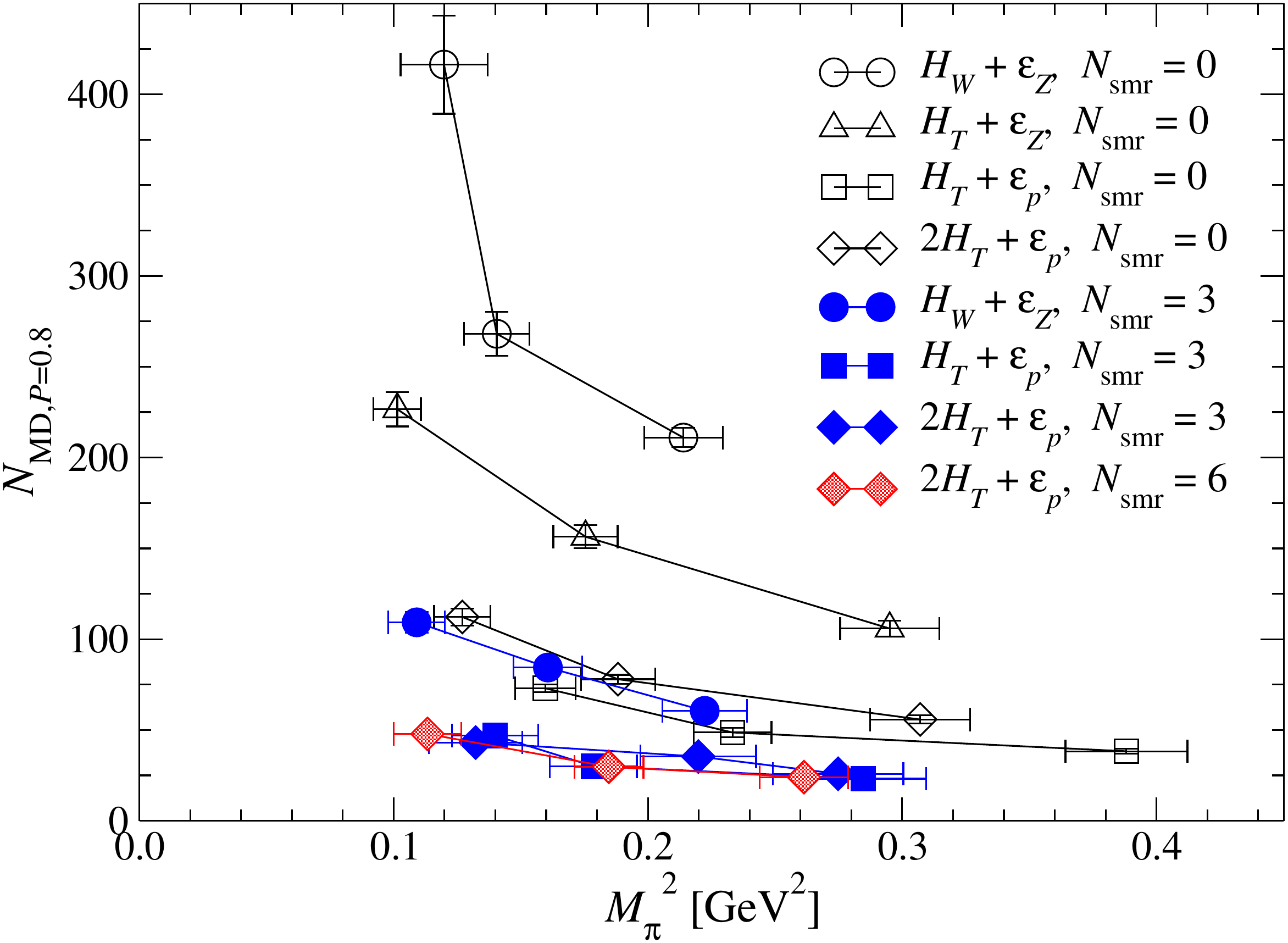}
   \hspace{3mm}
   \includegraphics[angle=0,width=0.49\linewidth,clip]%
                   {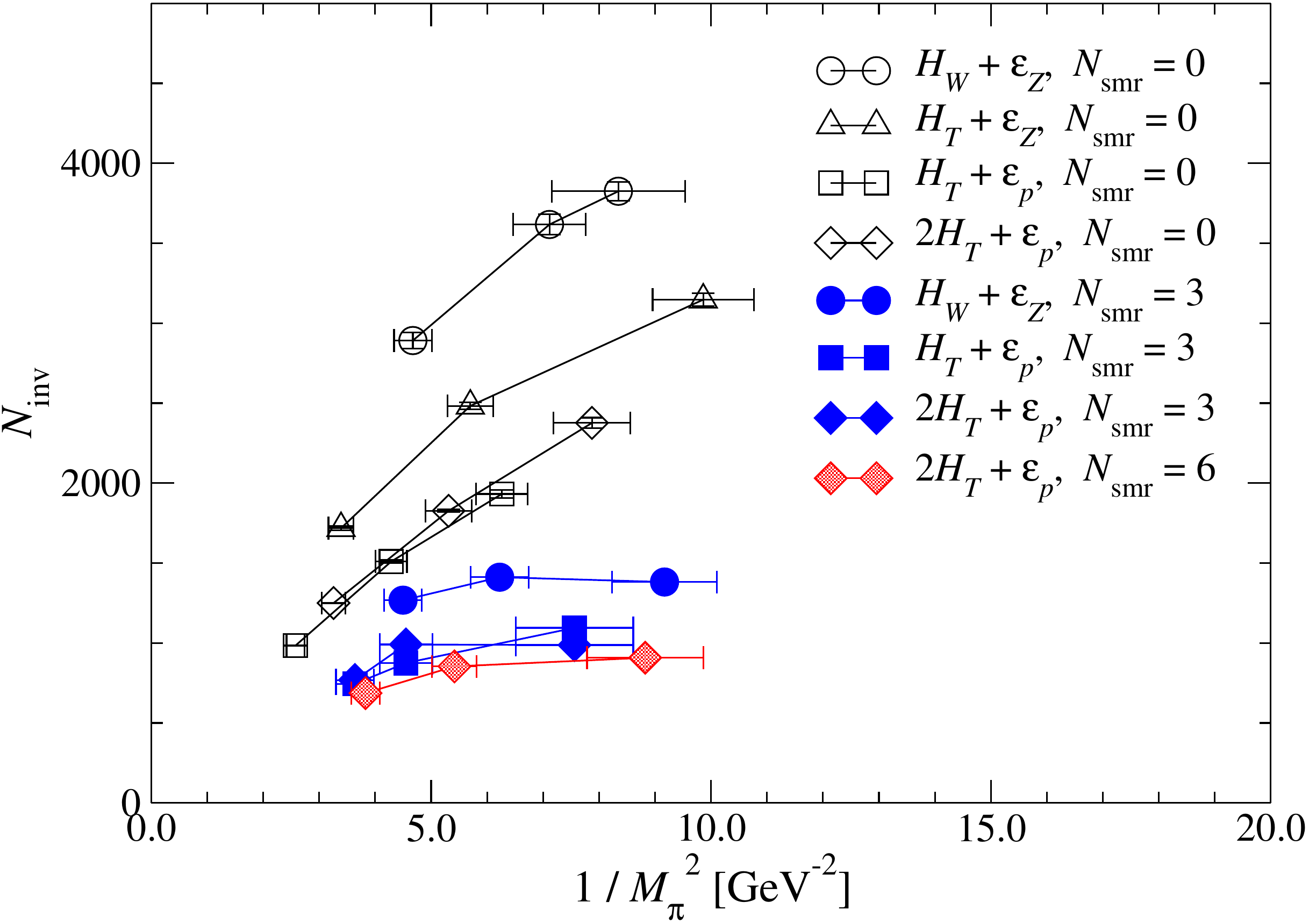}
   \vspace{-1mm}
   \caption{
     Left panel: 
     number of MD steps $\Nmdref$
     to attain 80\,\% acceptance rate. 
     Data for different formulations are plotted 
     in different symbols as a function of $M_\pi^2$.
     Right panel: 
     CG iteration count $\Ninv$ as a function of $M_\pi^{-2}$.
   }
   \label{fig:compara:NMD_and_Ninv}
\end{center}
\vspace{-5mm}
\end{figure}

\begin{figure}[t]
\begin{center}
   \includegraphics[angle=0,width=0.48\linewidth,clip]%
                   {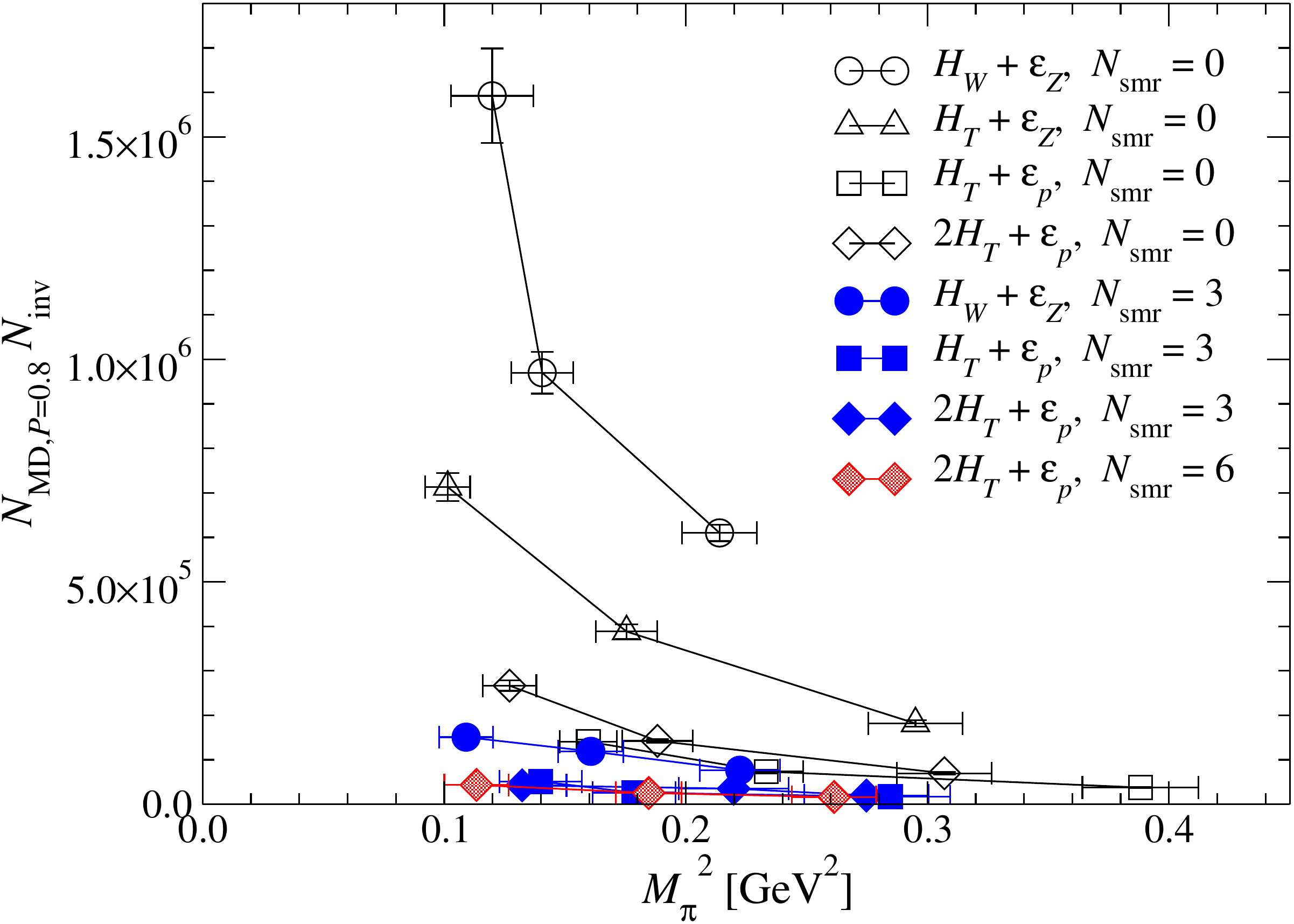}
   \hspace{3mm}
   \includegraphics[angle=0,width=0.48\linewidth,clip]%
                   {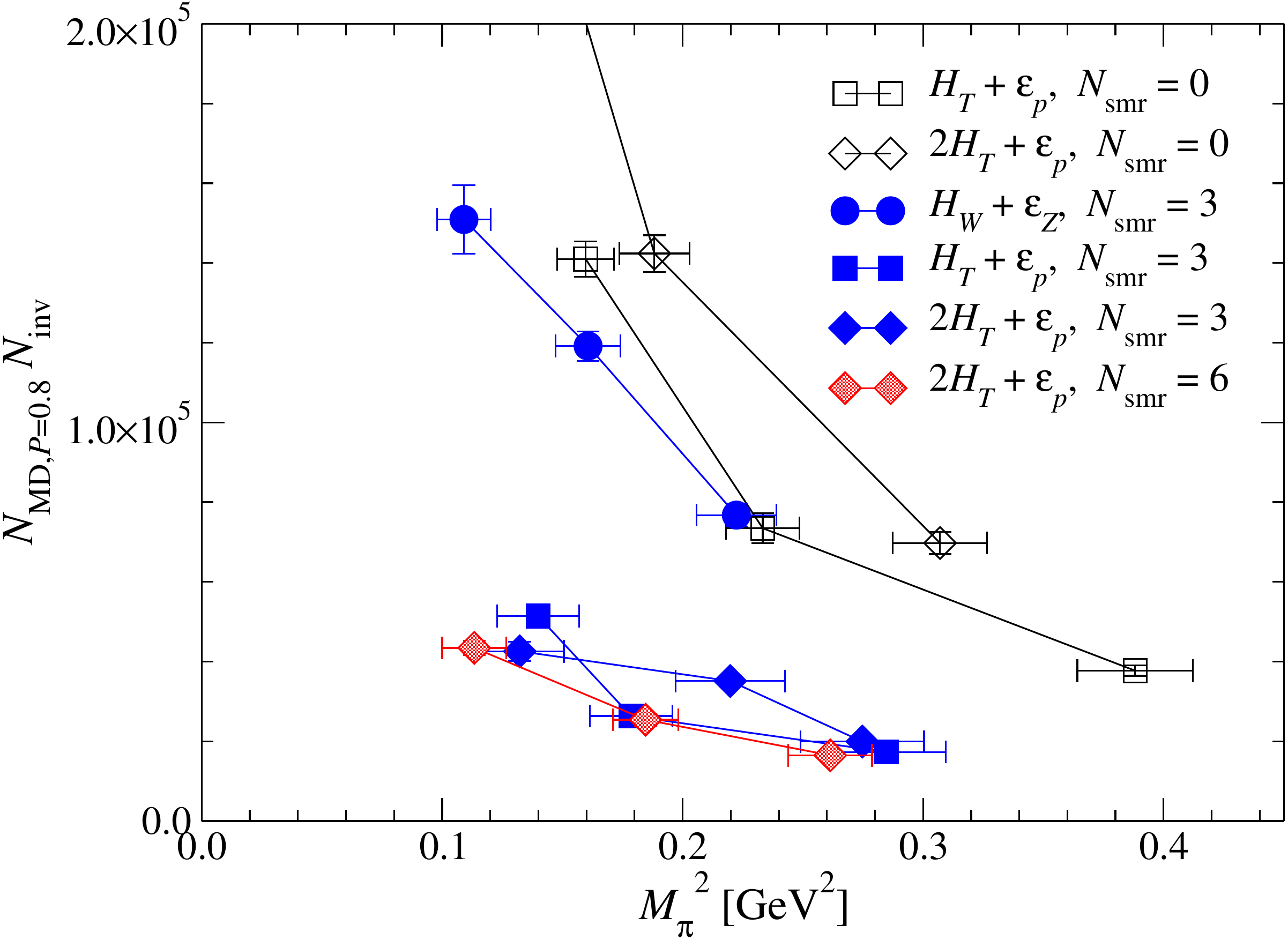}
   \vspace{-1mm}
   \caption{
     A measure of CPU cost per HMC trajectory $\Nmdref\,\Ninv$. 
     The left panel shows all data, 
     whereas the right panel is an enlargement
     of a region of small $\Nmdref\,\Ninv$
     to focus on computationally cheaper formulations. 
   }
   \label{fig:compara:NMDxNinv}
\end{center}
\vspace{-6mm}
\end{figure}

In each simulation,
we keep the acceptance rate of $P\!\simeq\!0.7$\,--\,0.9
using a moderately small step size $\dtau$ 
for the Molecular Dynamics (MD) integration.
The number of the  MD steps to attain a reference value $P\!=\!0.8$, 
which is denoted by $\Nmdref$ in the following,
is estimated from the relations holding at small $\dtau$ 
\bea
   P = \mbox{erfc}\left( \frac{1}{2}\sqrt{\dHave} \right), 
   \hspace{5mm}
   \dHave \propto \dtau^4,
\eea
where $\dHave$ represents the Monte Carlo average of 
the change of the Hamiltonian 
due to the discretized MD integration.
Figure~\ref{fig:compara:NMD_and_Ninv}
compares $\Nmdref$ and the iteration count for CG
per MD step, denoted by $\Ninv$,
among the tested formulations. 
We observe that these two measures of the CPU cost 
significantly decrease by 
i) switching from $H_W$ to $(2)H_T$,
ii) switching from $\sgnZ$ to $\sgnP$, 
and iii) applying smearing $(\Nsmr\!\geq\!3)$.
On the other hand,
there is no large difference in these measures
between Shamir-type kernels
($H_T$ and $2H_T$) and between $\Nsmr\!=\!3$ and 6.

The product $\Nmdref\,\Ninv$ can be considered as a measure of the 
CPU cost per HMC trajectory. 
As plotted in Fig.~\ref{fig:compara:NMDxNinv}, 
the overlap formulation, 
namely the combination of $H_W$ and $\sgnZ$, 
turns out to be computationally very demanding. 
We can achieve about a factor of 20 acceleration at $M_\pi\!\simeq\!400$~MeV:
a factor of 5 by using $(2)H_T$ and $\sgnP$,
and an additional factor of 4 by smearing.
We may expect even bigger gain at smaller quark masses.

\begin{figure}[t]
\begin{center}
   \includegraphics[angle=0,width=0.475\linewidth,clip]%
                   {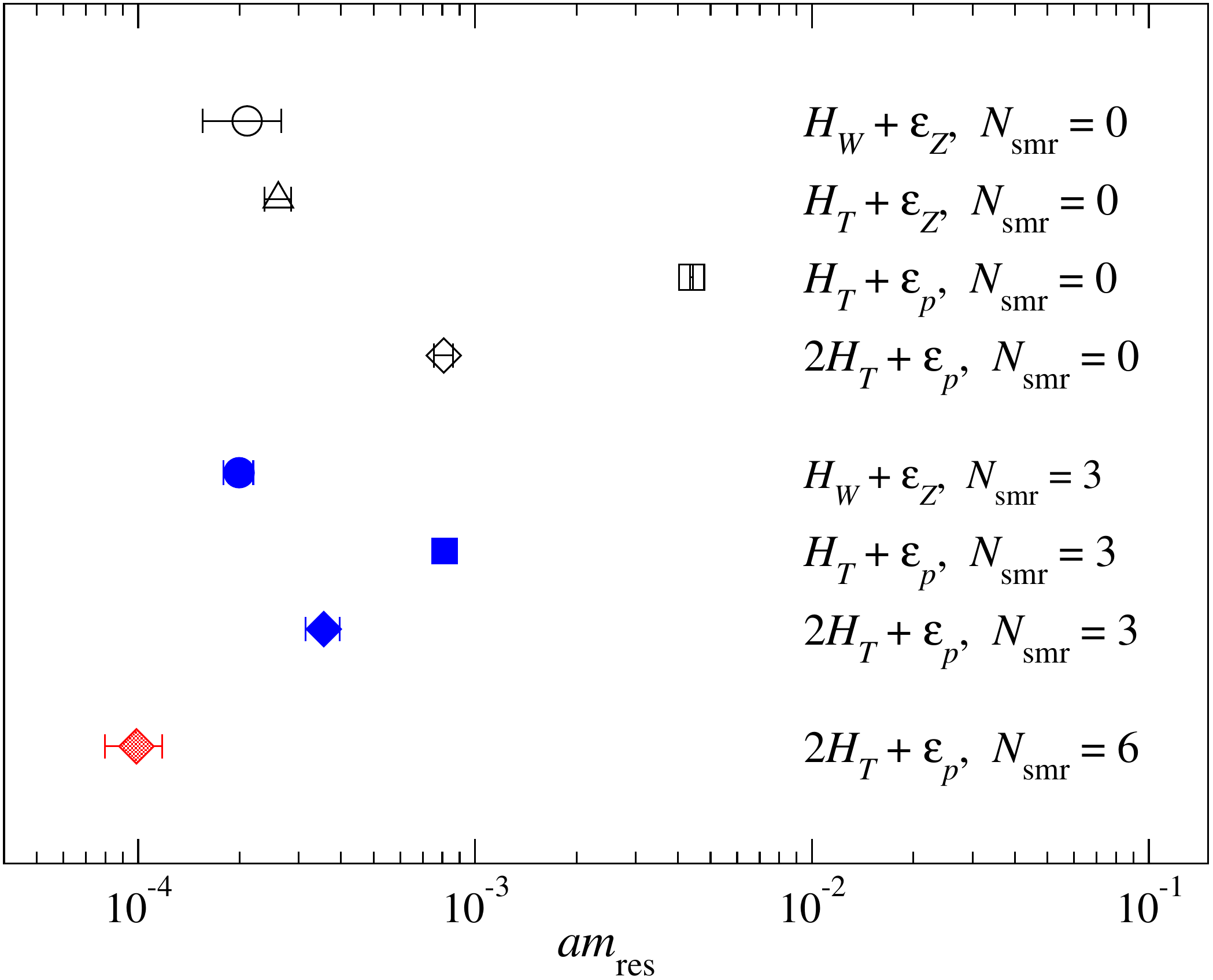}
   \hspace{5mm}
   \includegraphics[angle=0,width=0.465\linewidth,clip]%
                   {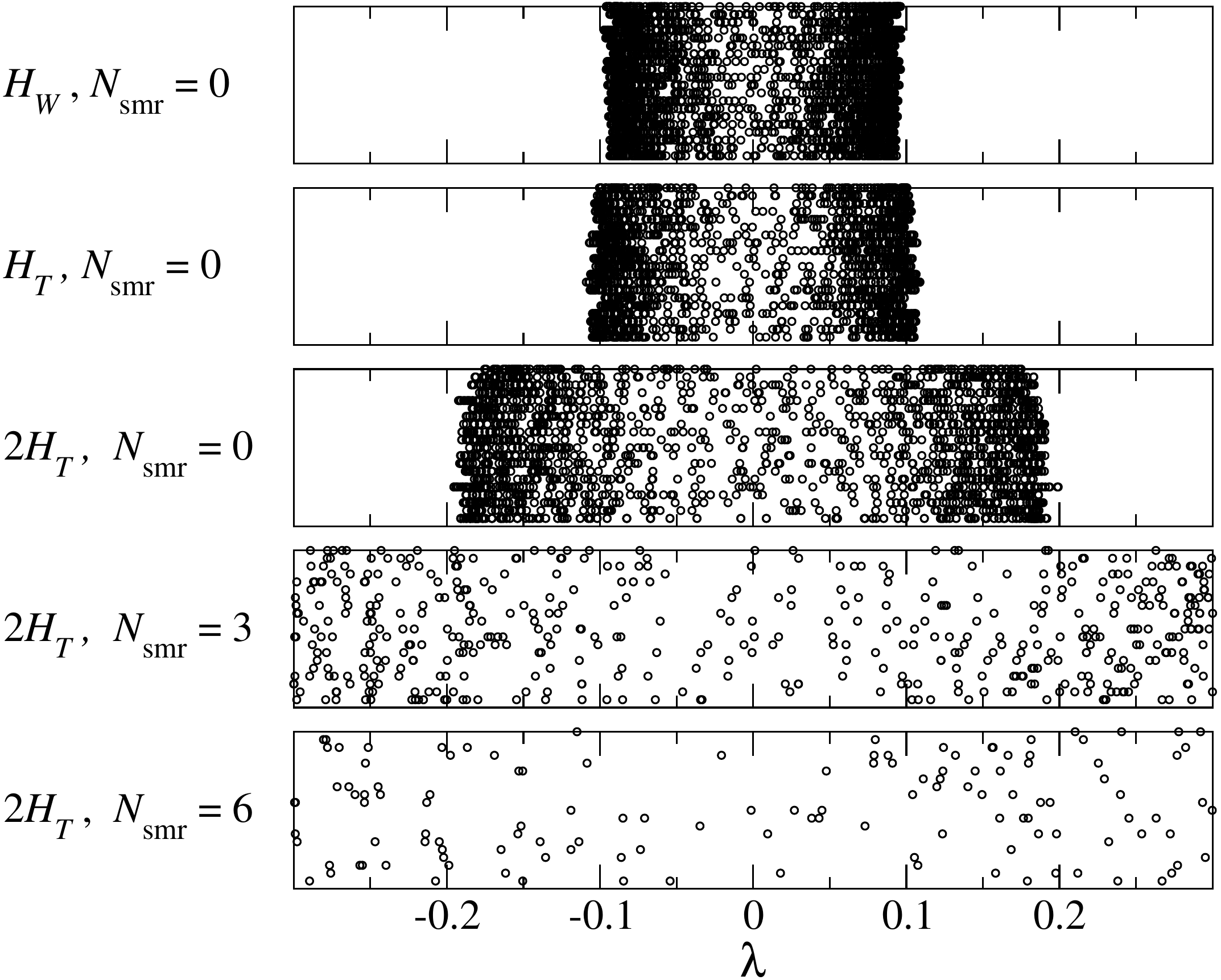}
   \vspace{-2mm}
   \caption{
     Left panel: 
     a comparison of bare residual quark mass in lattice units,
     $am_{\rm res}$.
     Right panels: 
     distribution of eigenvalue $\lambda$ 
     for several choices of the kernel operator.
     Note that we only plot the lowest 100\,--\,150 eigenvalues
     for thin-link kernels $(\Nsmr\!=\!0)$.
   }
   \label{fig:compara:m_res}
\end{center}
\vspace{-3mm}
\end{figure}

\begin{figure}[t]
\begin{center}

   \includegraphics[angle=0,width=0.47\linewidth,clip]%
                   {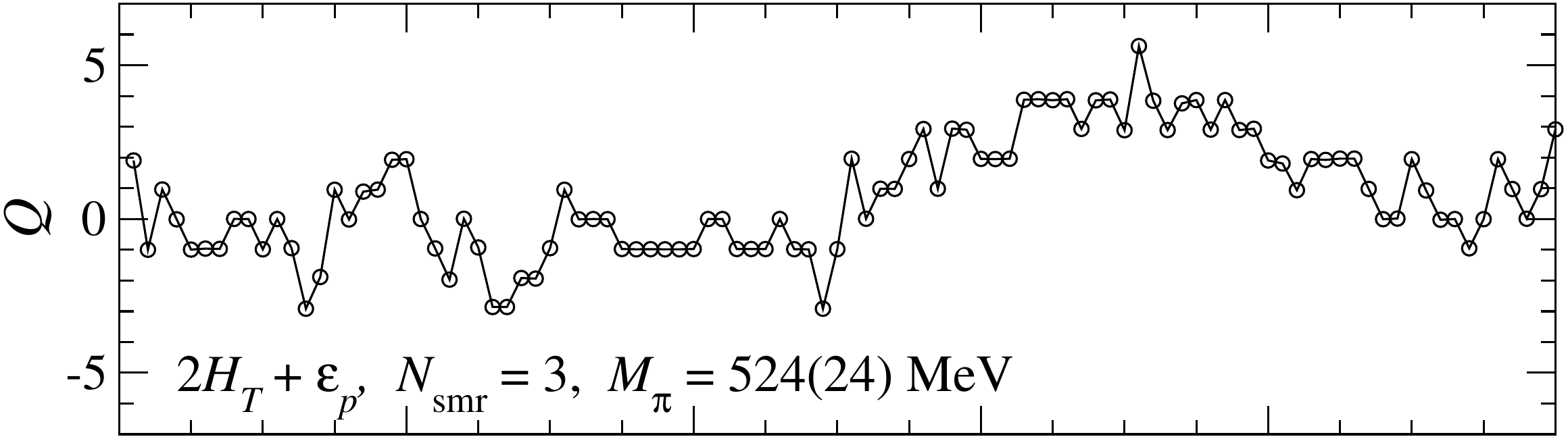}
   \hspace{2mm}
   \includegraphics[angle=0,width=0.47\linewidth,clip]%
                   {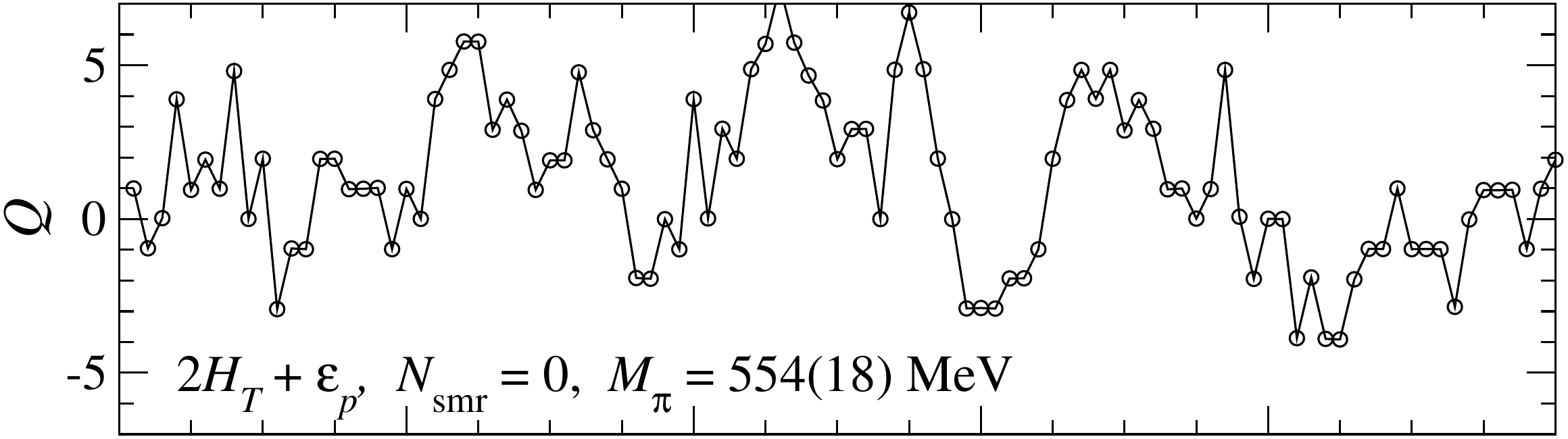}
   \vspace{4mm}

   \includegraphics[angle=0,width=0.48\linewidth,clip]%
                   {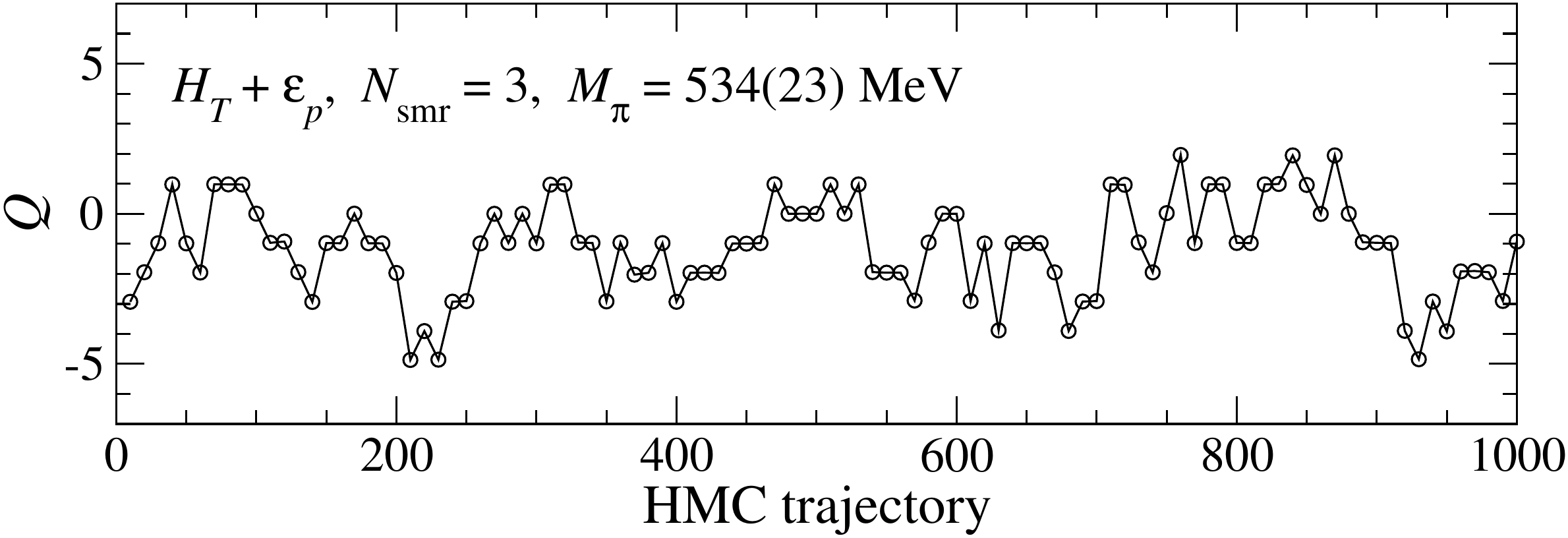}
   \hspace{2mm}
   \includegraphics[angle=0,width=0.48\linewidth,clip]%
                   {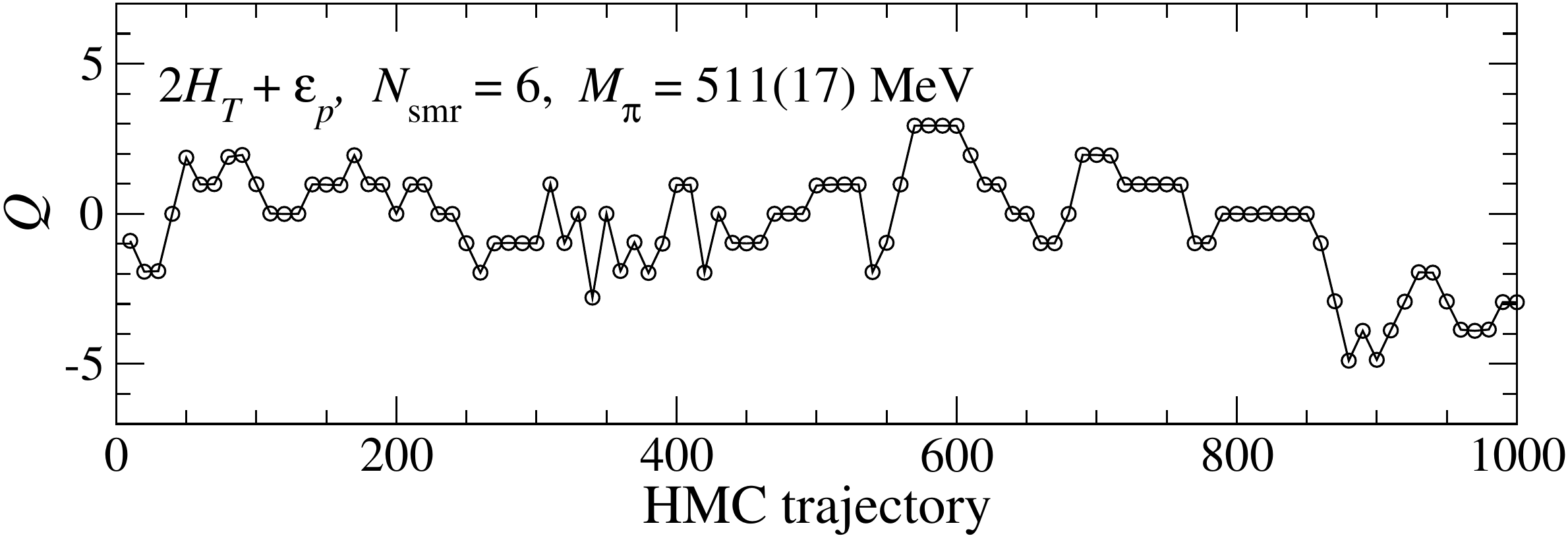}
   \vspace{-2mm}
   \caption{
     Monte Carlo history of topological charge.
     Left panels compare data for the Shamir-type kernels 
     ($2H_T$ and $H_T$) with $\sgnP$ and $\Nsmr\!=\!3$,
     whereas right panels show data for $2H_T$
     with different $\Nsmr$ (0 and 6).
   }
   \label{fig:compara:topology}
\end{center}
\vspace{-6mm}
\end{figure}

These computationally cheaper formulations are, however,
off from practical use, 
if they largely violate chiral symmetry.
We compare residual quark mass $\mres$ in Fig.~\ref{fig:compara:m_res}.
Since the min-max approximation can satisfy 
$|\sgnZ(x)|^2 \!\sim\! 1$ in its approximation range, 
$\sgnZ$ leads to the least $\mres$ at a given $\Nsmr$.
With our choice of $N_5\!=\!12$, however, 
$|\sgnP(x)|^2$ largely deviates from unity 
at $x \lesssim 0.3$,
where thin-link kernels have many low-lying modes
as shown in Fig.~\ref{fig:compara:m_res}.
Scaling of the kernel $(H_T \!\to\! 2H_T)$ and smearing $(\Nsmr\!=\!3)$ 
are very effective to suppress these low-lying modes
leading to an order of magnitude smaller $\mres$
compared to the standard domain-wall fermions.
Larger $\Nsmr$ is better in reducing $\mres$
but may distort short distance physics. 
We refer to Ref.~\cite{JLQCD:mres} for more detailed discussions.

Figure~\ref{fig:compara:topology} shows examples of 
the Monte Carlo history of the topological charge $Q$. 
A low-lying eigenvalue flips its sign 
along a tunneling between topological sectors.
While scaling and smearing suppress the low-lying modes,
the comparison in Fig.~\ref{fig:compara:topology} suggests that 
these techniques do not prevent the topological tunneling 
at $a^{-1}\!\simeq\!2$~GeV.

From this comparative study,
we conclude that the combination of $2H_T$ and $\sgnP$
with $\Nsmr\!=\!3$ is the best choice 
among the tested formulations.


\begin{table}[t]
\begin{center}
\caption{
   Status of our simulations at $a^{-1}\!\simeq\!2.4$~GeV.
   The third column shows the choice of the MD integrator,
   namely the leap-frog (LF) or Omelyan (O) integrator.
   We also list time per HMC trajectory
   on the whole machine of BlueGene/Q at KEK
   in the last column.
}
\vspace{2mm}
\label{tbl:prod:simu:2.4GeV}
\begin{tabular}{ll|lllllllllll}   \hline
   $am_{ud}$  & $am_s$  & MD        & $N_{\rm MD}$  
             & traj     
             & $P$     & $\dHave$  & $\edHave$ 
             & min/traj
   \\ \hline \hline
   0.019 & 0.040 & LF  & 10 &
           3000  & 0.78(1)  & 0.19(1)  & 0.99(1) & 2.7 
   \\ \hline 
   0.012 & 0.040 & LF  & 13 & 
           2000  & 0.78(1)  & 0.17(1)  & 1.00(1) & 3.5
   \\ 
   0.012 & 0.040 & O   & 3  & 
           1000  & 0.89(1)  & 0.07(2)  & 1.01(1) & 2.0
   \\ \hline
   0.007 & 0.040 & LF  & 16 & 
           1000  & 0.74(1)  & 0.23(2)  & 1.04(3) & 4.4
   \\ 
   0.007 & 0.040 & O   & 4  & 
           2000  & 0.90(1)  & 0.06(1)  & 1.00(1) & 2.6
   \\ \hline \hline
   0.019 & 0.030 & LF  & 10 & 
           3000  & 0.79(1)  & 0.17(1)  & 1.00(1) & 2.8
   \\ \hline
   0.012 & 0.030 & LF  & 13 & 
           2000  & 0.79(1)  & 0.14(1)  & 1.02(2) & 3.6
   \\
   0.012 & 0.030 & O   & 3  & 
           1000  & 0.88(1)  & 0.10(3)  & 1.00(2) & 2.0
   \\ \hline
   0.007 & 0.030 & LF  & 16 & 
           2000  & 0.72(1)  & 0.27(2)  & 1.00(2) & 4.5
   \\
   0.007 & 0.030 & O   & 4  & 
           1000  & 0.89(1)  & 0.08(2)  & 0.99(1) & 2.6
   \\ \hline
\end{tabular}
\end{center}
\vspace{-4mm}
\end{table}

\section{Large-scale simulations}

We have launched large-scale simulations of $N_f\!=\!2\!+\!1$ QCD
with good chiral symmetry,
namely with $\mres$ well below 
the physical up and down quark mass $m_{ud,\rm phys}$.
The tree-level Symanzik gauge action is combined with 
the fermion formulation chosen by the comparative study
to be consistent with our $O(a^2)$-improvement program
for heavy quark physics~\cite{JLQCD:Oa2-impr}.
For controlled continuum and chiral extrapolations, 
we are planning to simulate the pion masses of 
500, 400, 300~MeV (and even smaller)
at four values of the lattice cut-off 
$a^{-1}\!\simeq\!2.4$, 3.0, 3.6 and 4.8~GeV.
Finite volume effects are suppressed to 1\,--\,2\,\% level 
by keeping $M_\pi L \!\gtrsim\! 4$.
These simulations are being carried out on BlueGene/Q at KEK
(6 racks with a peak speed of 1.258 PFLOPS).

Table~\ref{tbl:prod:simu:2.4GeV} shows the current status 
of our simulations on a $32^3\!\times\!64\!\times\!12$ lattice 
at $\beta\!=\!4.17$, 
where $a^{-1}$ determined from $r_0$ is expected to be $\simeq2.4$~GeV.
The three values of the bare light quark mass $m_{ud}$ 
correspond to $M_\pi\!\approx\!500$, 400 and 300~MeV,
whereas we take two strange quark masses ($m_s$'s)
near its physical value $m_{s,\rm phys}$.
We employ the Hasenbusch preconditioning~\cite{multimass}
with the mass parameter $am^\prime\!=\!0.150$ for two degenerate light flavors,
and the rational HMC algorithm~\cite{RHMC} 
for the single strange flavor.
We had started our simulations with the simple leap-frog MD integrator,
which was later switched to the Omelyan integrator~\cite{Omelyan}
leading to a factor of 2 speed-up.
We keep reasonably high acceptance rate $P\!\simeq\!0.7$\,--\,0.9
and confirm that 
$\edHave\!=\!1$ derived from the area preserving property of HMC 
is well satisfied.

\begin{table}[t]
\begin{center}
\caption{
   Status of our simulations at $a^{-1}\!\simeq\!3.6$~GeV.
}
\vspace{2mm}
\label{tbl:prod:simu:3.6GeV}
\begin{tabular}{ll|lllllllllll}   \hline
   $am_{ud}$  & $am_s$  & $am^\prime$ & $N_{\rm MD}$  & traj     
             & $P_{\rm HMC}$  & $\dHave$    & min/traj
   \\ \hline \hline
   0.0120 & 0.0250 & 0.10 & 4 &
            430   & 0.84(2)  & 0.10(2)  & 3.6
   \\ \hline 
   0.0080 & 0.0250 & 0.08 & 4 &
            330   & 0.85(2)  & 0.06(2)  & 4.2
   \\ \hline 
   0.0042 & 0.0250 & 0.04 & 4 &
            235   & 0.92(3)  & 0.04(2)  & 5.9
   \\ \hline\hline
   0.0120 & 0.0180 & 0.10 & 4 &
            --    & --       & --       & --       
   \\ \hline
   0.0080 & 0.0180 & 0.08 & 4 &
            260   & 0.86(1)  & 0.05(1)  & 4.3
   \\ \hline
   0.0042 & 0.0180 & 0.04 & 4 & 
            280   & 0.86(3)  & 0.02(2)  & 6.0
   \\ \hline
\end{tabular}
\end{center}
\vspace{-4mm}
\end{table}

We are also carrying out simulations at a larger lattice cut-off 
$a^{-1}\!\simeq\! 3.6$~GeV ($\beta\!=\!4.35$) 
on $48^3\!\times\!96\!\times8$.
The current status is summarized in Table~\ref{tbl:prod:simu:3.6GeV}. 
We increase the unit trajectory length to $\tau\!=\!2$ 
based on our preparatory study on the auto-correlation (see below).
Our choice of the fermion action as well as 
careful tuning of $m^\prime$ at each $m_{ud}$ 
enable us to achieve the high acceptance rate $P\!\gtrsim\!0.85$ 
with small $\Nmd\!=\!4$.
We expect half a year to accumulate 10,000 MD time
on this large volume by using BlueGene/Q at KEK.
This will be accelerated 
by further optimization of our simulation code~\cite{JLQCD:code}.

\FIGURE{
   \label{fig:prod:m_res}
   \includegraphics[angle=0,width=0.48\linewidth,clip]%
                   {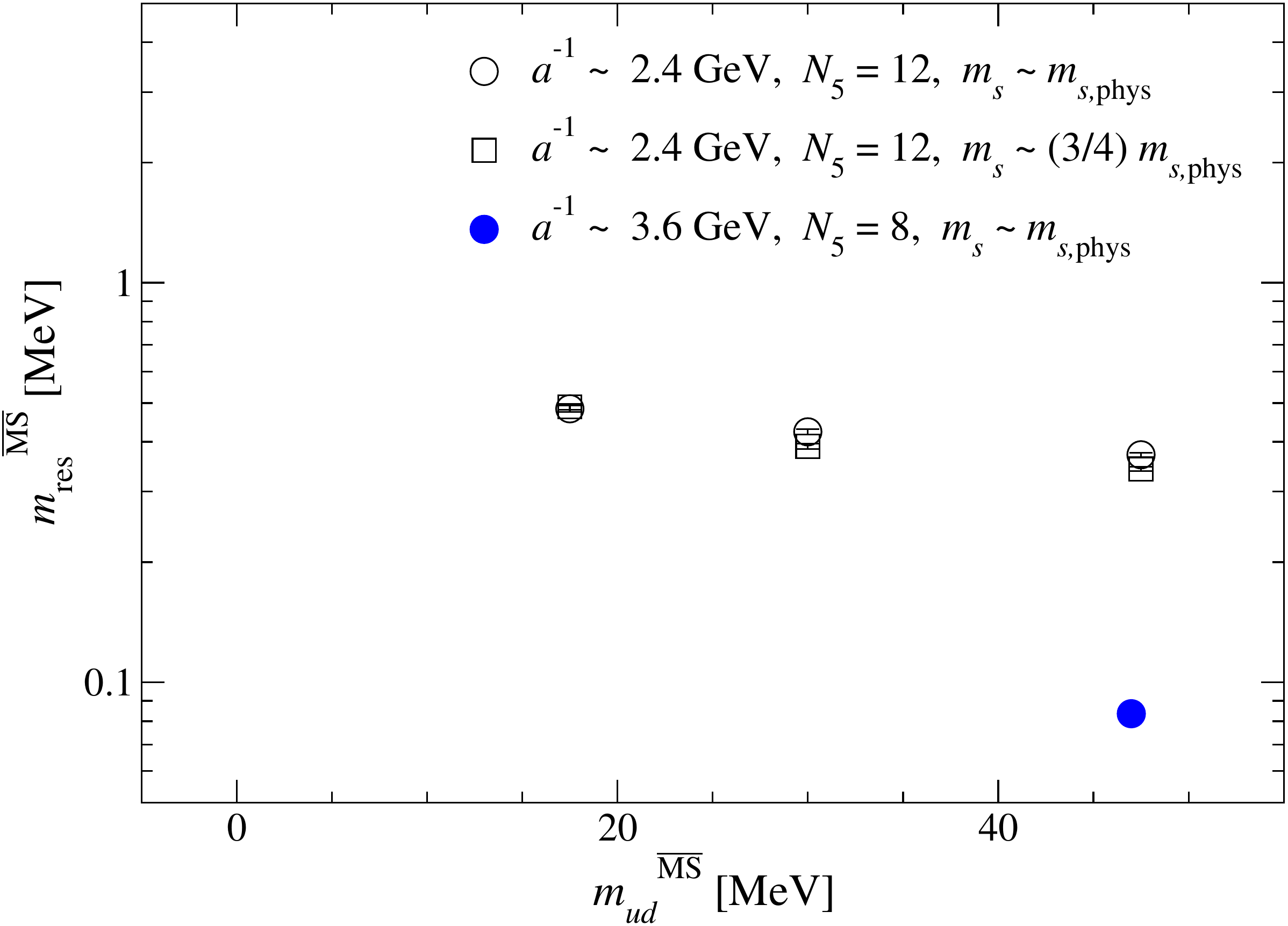}
   \vspace{-3mm}
   \caption{
      $m_{\rm res}$ in $\overline{\rm MS}$ scheme at 2~GeV.
   }
}
We plot $\mres$ from these simulations in Fig.~\ref{fig:prod:m_res},
where the renormalization factor to 
the $\overline{\rm MS}$ scheme at 2~GeV
is roughly estimated by matching 
our estimate of the bare value of $m_{s,\rm phys}$ 
with a world average~\cite{FLAG} in that scheme.
It turns out that 
$\mres\!\simeq\!0.5$~MeV
at $a^{-1}\!\simeq\!2.4$~GeV with $N_5\!=\!12$. 
At $a^{-1}\!\simeq\!3.6$~GeV,
$\mres$ is even smaller $(\simeq\!0.1\mbox{MeV})$
with smaller $N_5\!=\!8$.
While these $\mres$'s are already much smaller than $m_{ud,\rm phys}$,
we are considering to further reduce $\mres$ 
by reweighting~\cite{JLQCD:reweight}.

\begin{figure}[t]
\begin{center}
   \includegraphics[angle=0,width=0.48\linewidth,clip]%
                   {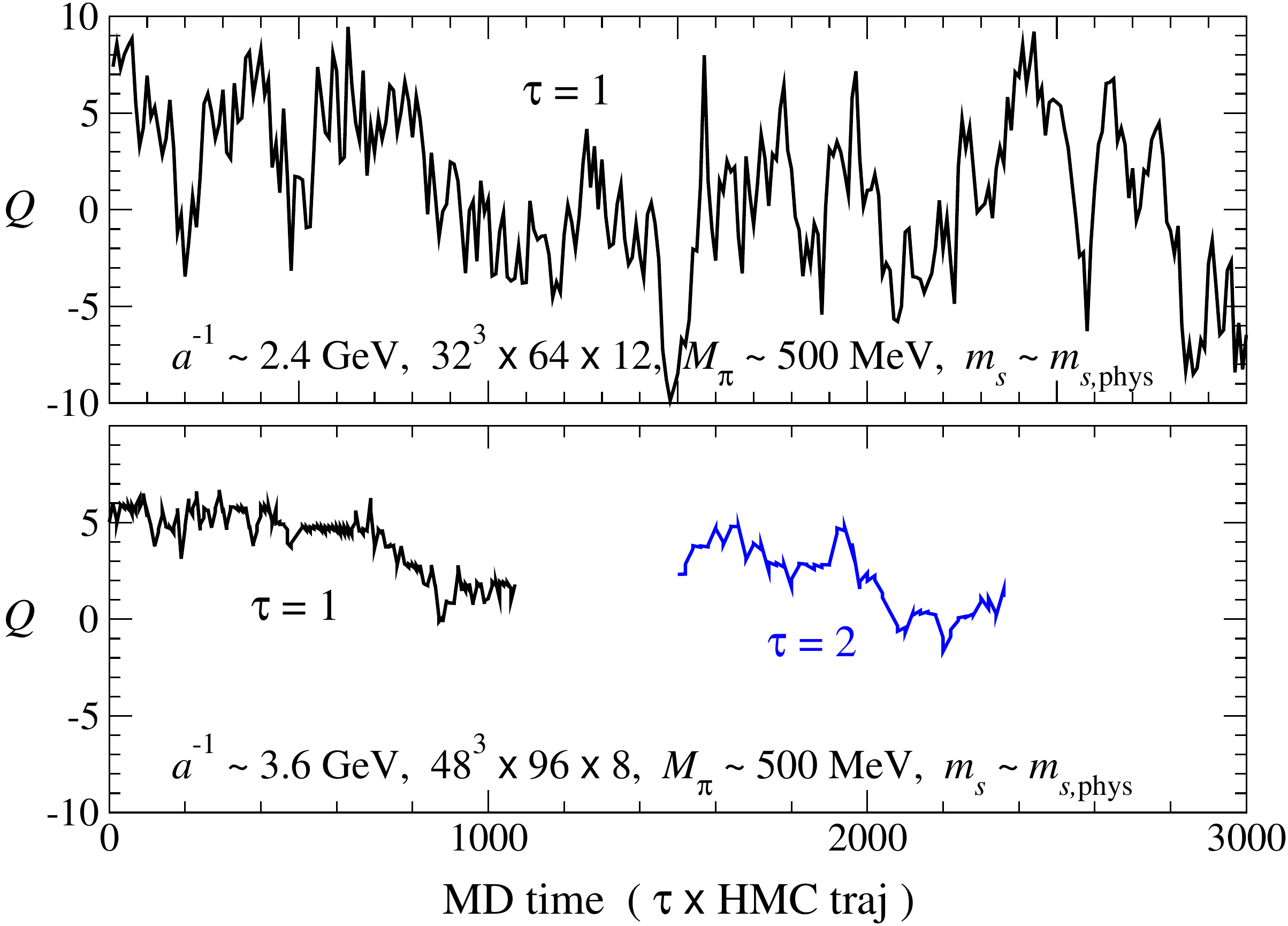}
   \hspace{3mm}
   \includegraphics[angle=0,width=0.48\linewidth,clip]%
                   {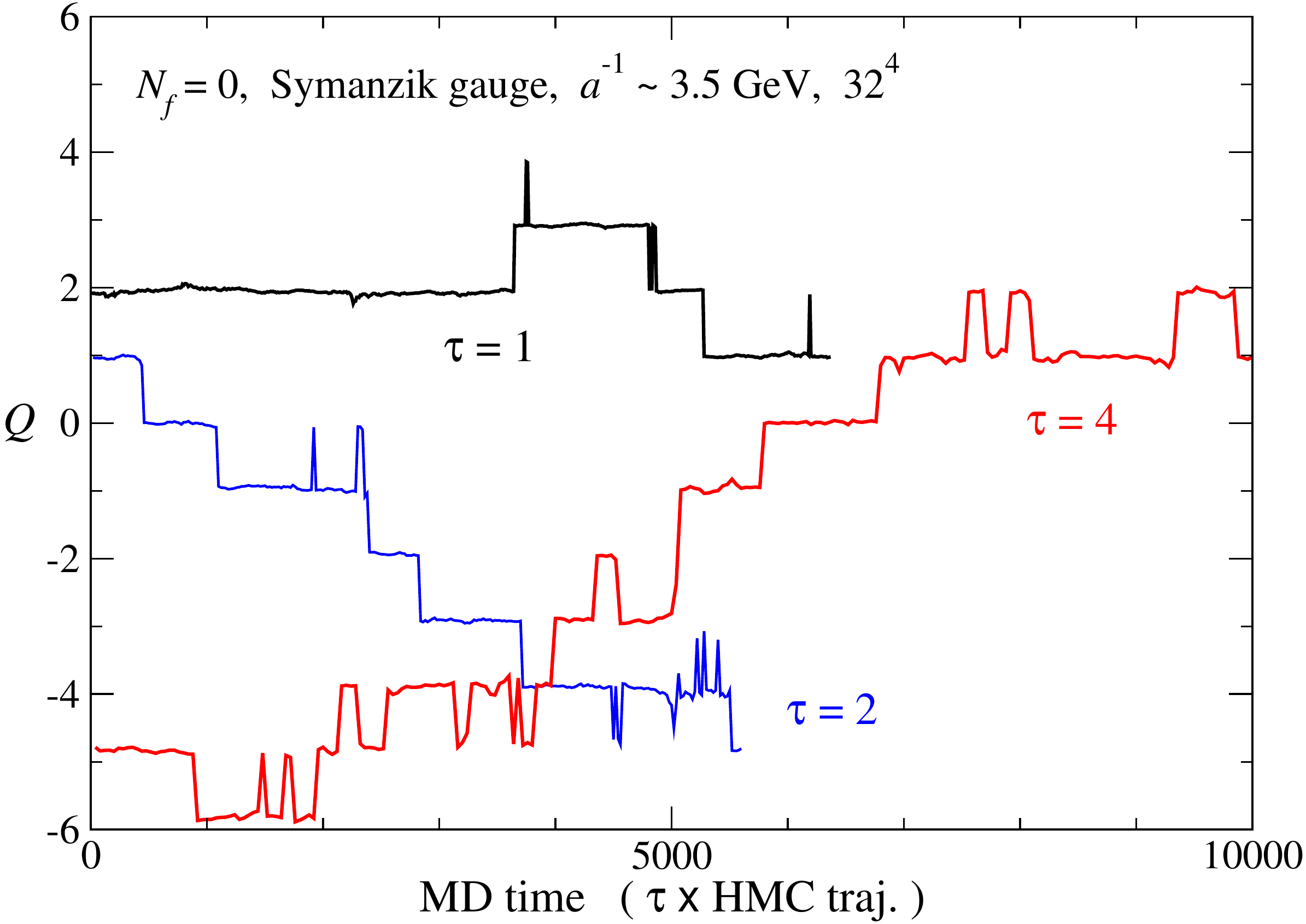}
   \vspace{-1mm}
   \caption{
     Left panels: 
     Monte Carlo history of topological charge $Q$ 
     in our simulations of $N_f\!=\!2+1$ QCD 
     at $a^{-1}\!\simeq\!2.4$ (left-top panel) 
     and 3.6~GeV (left-bottom panel).
     Right panel: 
     history of $Q$ in our study in quenched QCD 
     at $a^{-1}\!\simeq\!3.5$~GeV
     with different values of $\tau$.
   }
   \label{fig:prod:auto-corr}
\end{center}
\vspace{-5mm}
\end{figure}

In Fig.~\ref{fig:prod:auto-corr},
we compare the topological tunneling 
at $a^{-1}\!\sim\!2.4$ and 3.6~GeV.
The auto-correlation largely increases 
by approaching the continuum limit
with the unit trajectory length $\tau$ held fixed.
As suggested in Ref.~\cite{large_tau},
we observe that topology changes more frequently 
with larger $\tau$ 
in our study in quenched QCD 
at a similar cut-off $a^{-1}\!\simeq\!3.5$~GeV.
This observation leads us to
increase $\tau$ when exploring $a^{-1}$ above 2.4~GeV
to accelerate our Monte Carlo sampling of topological sectors.

\vspace{3mm}

In this article,
we reported on our new project of large-scale simulations 
of $N_f\!=\!2\!+\!1$ QCD with good chiral symmetry. 
The lattice action is chosen by the comparative study 
to reduce $\mres$ well below the physical quark masses 
and achieve a factor of 20 acceleration 
compared to the overlap formulation.
We are planning to accumulate high statistics of 10,000 MD time
for a precision study of QCD.
Our preliminary results on the light hadron physics 
were presented at this conference~\cite{JLQCD:spectrum}.

\vspace{3mm}

Numerical simulations are performed on Hitachi SR16000 and 
IBM System Blue Gene Solution 
at High Energy Accelerator Research Organization (KEK) 
under a support of its Large Scale Simulation Program (No.~12/13-04).
This work is supported in part by 
the Grants-in-Aid for Scientific Research 
(No.~21674002, 25287046), 
the Grant-in-Aid for Scientific Research on Innovative Areas
(No. 2004: 20105001, 20105002, 20105003, 20105005, 23105710),
and SPIRE (Strategic Program for Innovative Research).
\vspace{0mm}


\vspace{-3mm}

\end{document}